\newcommand{\gev}{\, {\rm GeV}}

\newcommand{\beq}{\begin{equation}}
\newcommand{\eeq}{\end{equation}}
\newcommand{\bea}{\begin{eqnarray}}
\newcommand{\eea}{\end{eqnarray}}

\newcommand{\gsim}{\lower.7ex\hbox{$\;\stackrel{\textstyle>}{\sim}\;$}}
\newcommand{\lsim}{\lower.7ex\hbox{$\;\stackrel{\textstyle<}{\sim}\;$}}

\documentclass[aps,prev,twocolumn,preprintnumbers,%
               floatfix,nofootinbib]{revtex4}
\usepackage{graphicx}
\usepackage{mathrsfs}
\usepackage{amssymb}
\usepackage{verbatim}

\def\stacksymbols #1#2#3#4{\def\theguybelow{#2}
    \def\vp{\lower#3pt}
    \def\sp{\baselineskip0pt\lineskip#4pt}
    \mathrel{\mathpalette\intermediary#1}}

\def\intermediary#1#2{\vp\vbox{\sp
     \everycr={}\tabskip0pt
     \halign{$\mathsurround0pt#1\hfil##\hfil$\crcr#2\crcr
              \theguybelow\crcr}}}

\def\comment#1{}
\def\to{\rightarrow}

\def\u1x{U(1)_X}
\newcommand{\nc}{\newcommand}
\nc{\LL}{L}
\nc{\vv}{\tilde{v}}
\nc{\ccdot}{\!\cdot\!}
\nc{\gsm}{G_{SM}}
\nc{\vfive}{\mathbf{5}\oplus\mathbf{\overline{5}}}
\nc{\vten}{\mathbf{10}\oplus\mathbf{\overline{10}}}
\nc{\zhol}{Z^{\rm hol}}
\nc{\xfb}{\,{\rm fb}}

\begin{document}

%
%

\preprint{MCTP-06-35}



\title{Model-Independent Description and Large Hadron Collider\\
Implications of Suppressed Two-Photon Decay of a Light Higgs Boson }

\author{Daniel Phalen, Brooks Thomas, James D. Wells}
\vspace{0.2cm}
\affiliation{
Michigan Center for Theoretical Physics (MCTP) \\
Department of Physics, University of Michigan, Ann Arbor, MI 48109}

\begin{abstract}
For a Standard Model Higgs boson with mass between 115 GeV and 150 GeV, the two-photon
decay mode is important for discovery at the Large Hadron Collider (LHC). We describe the
interactions of a light Higgs boson in a more model-independent fashion, and consider the
parameter space where there is no two-photon decay mode.  We argue from generalities that
analysis  of the $t\bar t h$ discovery mode outside its normally thought of range of
applicability is especially needed under these circumstances.  We demonstrate the general
conclusion with a specific example of parameters of a type I two-Higgs doublet theory,
motivated by ideas in strongly coupled model building.  We then specify a complete set of
branching fractions and discuss the implications for the LHC.

\end{abstract}

\maketitle


\maketitle


\setcounter{equation}{0}



{\it Introduction.} The Standard Model (SM) explanation for Electroweak Symmetry breaking
(EWSB) -- that one condensing Higgs boson doublet carries all burdens of elementary
particle mass generation -- remains speculative.  Experimental results are consistent
with this simple explanation, albeit for an increasingly shrinking window of Higgs boson
mass. Precision electroweak data suggest that the Higgs boson must be less than $\sim
200\gev$~\cite{LEPEWWG}, otherwise the virtual effects of the Higgs
boson cause theory predictions to be incompatible with the data.
Direct searches at LEP~II indicate that the Higgs
boson mass must be greater than $114\gev$~\cite{:2001xx}.

Despite these passive successes of the SM Higgs boson, well-known theoretical concerns
involving naturalness and the hierarchy problem suggest that the simple SM explanation
is not complete. There have been an extraordinary number of interesting ideas that
have surfaced for a more palatable description of EWSB and elementary particle mass
generation, and each of them affects Higgs boson collider observables.  The effect could
be small, as in supersymmetry with very heavy superpartners and a SM-like
lightest Higgs boson, or large, as in strongly coupled theories of EWSB~\cite{Hill:2002ap}.

In this letter, we examine an enlargement of the Higgs sector. It resembles the Standard
Model in that it contains one light Higgs boson, but differs from the SM in the strengths
of this particle's interactions with other fields. There is a compelling reason for doing
this sort of analysis: soon data from the LHC, a \(pp\) collider with
\(\sqrt{s}=14\)~TeV, will teach us a great deal more about the theoretical nuances of
EWSB.  We begin our analysis by introducing a basic, model-independent parameterization
of the couplings of the light Higgs boson to SM particles.  We then examine how
modifications to these couplings can dramatically affect the collider observables most
relevant to Higgs searches.  In particular, we focus on the process \(gg\to h\to
\gamma\gamma\), which is one of the most promising channels for the discovery of a light
Higgs boson at the LHC, and demonstrate how such modifications to the Higgs couplings
affect the potential for discovery.


{\it Shutting off two-photon decays.} Higgs boson decays to two photons provide one of
the most important detection channels for a SM Higgs boson in the mass range
\(115\mathrm{~GeV}\lesssim m_h\lesssim 150\)~GeV.  Here, we wish to discuss the
possibility that the two-photon decay branching fraction of the Higgs boson is
significantly reduced compared to that of the SM.  This can occur for many reasons.  In
certain regions of parameter space within supersymmetry, for example, this is possible
when the partial width to the dominant mode, such as $h\to b\bar b$, is significantly
enhanced such that $B(h\to \gamma\gamma)= \Gamma(h\to\gamma\gamma)/\Gamma_{tot}\to 0$
because $\Gamma_{tot}$ is so large~\cite{Kane:1995ek}.  However, a rather extreme shift
in couplings is needed in general to suppress the $h\to \gamma\gamma$ branching fraction
to an insignificant level this way. A less extreme way that nature could reduce the
$h\to\gamma\gamma$ branching fraction is by simultaneously altering the various couplings
that enter the effective $h\gamma\gamma$ vertex such that a cancelation occurs.  We will
focus primarily on this second possibility.

The SM Higgs boson observables at the LHC involve tree-level interactions of the Higgs
boson with $WW$, $ZZ$, and $f\bar f$. A model-independent parameterization of these
interactions suggests that we multiply each of the vertices by an $\eta$-factor of \bea
g^{\rm sm}_{\mathit{hWW}}\to \eta_W g^{\rm sm}_{\mathit{hWW}},~~ g^{\rm
sm}_{\mathit{hZZ}}\to\eta_Z g^{\rm sm}_{\mathit{hZZ}}, ~~ g^{\rm sm}_{\mathit{h\bar f
f}}\to \eta_f g^{\rm sm}_{\mathit{h\bar f f}} \nonumber \eea
The relevant
observables also rely crucially on the Higgs boson interacting at the loop level with
$gg$ and $\gamma\gamma$, and to a lesser degree of importance $\gamma Z$. These
interactions can be sensitive to new particles entering at one-loop order. Experimental
data presently have little bearing on the question of how large an effect the new
particles can have on the effective $hgg$ or $h\gamma\gamma$ vertices, and so we will
parameterize their effects through effective operators.  In general, deviations of the
$h\gamma\gamma$ and $hgg$ couplings arise two ways: deviations of
couplings of SM particles in the loops, and extra corrections due to exotic particles or
effects contributing to the effective interactions.  The former  can be
parameterized in terms of the $\eta$ coefficients described above; the latter can be
parameterized by introducing new variables $\delta_g$ and $\delta_\gamma$.  In the limit
that the top quark is much heavier than the other SM fermions, the resulting operators
are \beq \left( \delta_\gamma +\eta_WF_1(\tau_W)+\eta_t\frac{4}{3}F_{1/2}(\tau_t)\right)
\frac{h}{v}\frac{\alpha}{8\pi}F_{\mu\nu}F^{\mu\nu} \eeq and \beq
\left(\delta_g+\eta_tF_{1/2}(\tau_t)\right) \frac{h}{v}\frac{\alpha_s}{8\pi}
G^a_{\mu\nu}G^{a\mu\nu},\eeq where $v\simeq 246\gev$ (SM Higgs VEV),
$\tau_i=4m_i^2/m_h^2$, and \bea
F_{1/2}(\tau)&=& -2\tau[1+(1-\tau)f(\tau)] \\
F_1(\tau)& =& 2+3\tau+3\tau(2-\tau)f(\tau) \eea and \beq
f(\tau)=\left\{ \begin{array}{cc} {\rm arcsin}^2(1/\sqrt{\tau}) & \tau\geq 1 \\
-\frac{1}{4}\left[ \log (\eta_+/\eta_-) -i\pi\right]^2 & \tau<1
\end{array}\right.
\eeq with $\eta_\pm =(1\pm \sqrt{1-\tau})$~\cite{Gunion:1989we}.
One should note that $F_1(\tau_i)$ and
$F_{1/2}(\tau_i)$ can be complex if $m_h>2m_i$, as this corresponds to internal lines
going on shell.  Since any Higgs boson worth its name has the property $\eta_W\neq 0$,
one can express the condition under which the effective $h\gamma\gamma$ vertex vanishes
as
\begin{equation} \left( \frac{\eta_t}{\eta_W}\right) =
  -\frac{3}{4}\left(\frac{1}{F_{1/2}(\tau_t)}\left( \frac{\delta_\gamma}{\eta_W} \right) +
  \frac{F_1(\tau_W)}{F_{1/2}(\tau_t)}\right)
\end{equation} In fig.~\ref{eta
relations}, we plot the contours of $\Gamma(h\to \gamma\gamma)=0$ in the $\eta_t/\eta_W$
vs. $\delta_\gamma/\eta_W$ plane. The various lines of the plot correspond to various
values of $m_h$ within the range of Higgs boson masses that are considered applicable for
discovery through their $h\to\gamma\gamma$ decay channel.

\begin{figure}
\includegraphics[width=8.5cm]{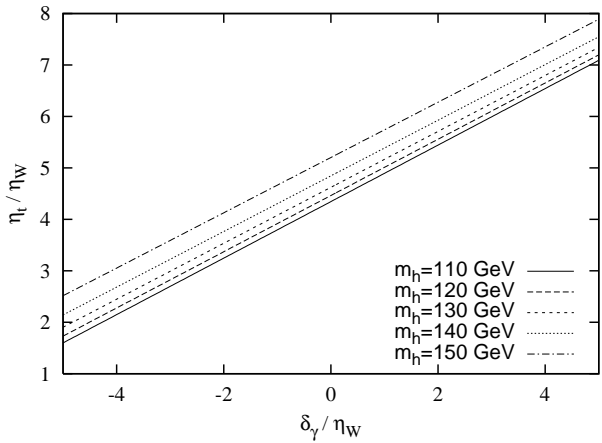}
\caption{Lines in  $\eta_t/\eta_W$ vs. $\delta_\gamma/\eta_W$ represent where
$\Gamma(h\to \gamma\gamma)\to 0$ for various values of $m_h$ in the range
of Higgs boson masses where the two-photon decays are thought to be important
for discovery at the LHC.}
\label{eta relations}
\end{figure}

{\it Theory discussion.} From experience in supersymmetry, exotic physics contributions
to $\delta_g$ or $\delta_\gamma$ can decouple very rapidly and have small
effects~\cite{Kane:1995ek}. Possible exceptions to this statement include a radion in
warped geometry that has large $\delta_g$ and $\delta_\gamma$ contributions due to
quantum breaking of conformal invariance~\cite{Giudice:2000av}, or
non-renormalizable operators with low-scale cutoff~\cite{Hall:1999fe}. If mixed with
a condensing Higgs boson, the lightest mass eigenstate can have significant $\delta_g$
and $\delta_\gamma$ contributions. Nevertheless, if we approximate that $\delta_g$ and
$\delta_\gamma$ are zero, we are left with the reasonable conclusion that the top quark
coupling to the Higgs boson is greatly enhanced compared to its SM value. This is to be
expected in the case where the $h\to \gamma\gamma$ decay rate goes to zero, since the
$W$-induced amplitude contribution is about six times larger than the top-quark induced
amplitude contribution and of the opposite sign. Thus, one expects that a theory with
reduction of the $W$ coupling and simultaneous increase in the top-quark coupling has the
potential to reduce and even zero out the $h\to \gamma\gamma$ amplitude.

Reducing $W$ couplings is easy: arrange for several states or mechanisms to contribute to
EWSB. Any individual Higgs boson state will have reduced couplings to $W$ since couplings
are proportional to the contribution that the Higgs boson
makes to the mass of the $W$ boson. As
for the coupling to the top quark, increasing its value is not only possible, but can be
expected when EWSB is shared among many sectors.  If the top quark couples to only one
Higgs boson \(H\) that does not fully generate the $W$ masses, then the Yukawa coupling
of the top quark to that Higgs has to have a larger value to generate the requisite top
quark mass (i.e., $m_t/\langle H\rangle > m_t/\langle H_{sm}\rangle$).

The above considerations lead us naturally to think in terms of a type I two-Higgs
doublet scenario~\cite{Haber:1978jt,Gunion:1989we}.  In such a scenario, both (complex)
doublets, which we denote \(\Phi_{f}\) and \(\Phi_{\mathit{EW}}\), obtain VEVs and
contribute to EWSB and vector boson mass generation, but only one of the Higgs bosons,
\(\Phi_{f}\), gives mass to the fermions.
Among the eight degrees of freedom in $\Phi_{EW}$ and $\Phi_f$ three are eaten by
$W^\pm_L,Z^0_L$, two become charged Higgses $H^\pm$, one becomes pseudo-scalar $A^0$, and
two become scalars $h^0,H^0$, where $h^0$ is the lightest.  Such a theory framework can
be motivated by the dynamics of a strongly coupled sector~\cite{Hill:2002ap}
which contributes to EWSB and
to the vector boson masses.  It is well-known that giving mass to the heavy fermions
simultaneously is difficult, and so an additional scalar (or effective scalar) can be
added to the theory that gives mass to the fermions (somewhat reminiscent of other
approaches~\cite{Simmons:1988fu}). Of course,
this second sector will necessarily contribute to vector boson masses as well (one cannot
hide EWSB from the vector bosons), and a complicated mixing among the two Higgs doublets
ensues.

In our study we define a mixing angle \(\beta\) between the two Higgs VEVs,
\(\tan\beta=\langle\Phi_f\rangle/\langle \Phi_{EW}\rangle\).  A second angle, \(\alpha\),
which parameterizes the mixing between the gauge and mass eigenstates of the CP-even
Higgs, is defined by the relation \beq \left(
\begin{array}{c} H^0 \\ h^0 \end{array}\right) =
\left( \begin{array}{cc} \cos\alpha & \sin\alpha \\
-\sin\alpha & \cos\alpha\end{array}\right) \left( \begin{array}{c}
\sqrt{2} {\rm Re}(\Phi_{EW}^0) \\
\sqrt{2} {\rm Re}(\Phi_f^0)
\end{array}\right)
\eeq The deviations in the couplings to SM fermions and vectors bosons of $h^0$ compared
to the SM Higgs depend only on these angles:
\beq
\eta_f=\frac{\cos\alpha}{\sin\beta},~~~\eta_{W,Z}=\sin(\beta-\alpha). \label{eq:etas}
\eeq


The value of $\alpha$ is computed from the full potential of the theory is
model-dependent, as are the masses of $H^\pm$ and $A^0$.  The most stringent experimental
constraint on these parameters comes from \(b\rightarrow s\gamma\). The experimental
limit is \(BR(b\rightarrow s\gamma)=(3.3\pm0.4)\times10^{-4}\) and in type~I Higgs
doublet models~\cite{Barger:1989fj},
\begin{equation}
  \Gamma(b\rightarrow s\gamma)=\frac{\alpha G_{F}^2 m_b^2}{128\pi^4}
  \left|A_{W}+\cot^2\beta A_{H}(m_{H^{\pm}})\right|^2,
  \label{bsg eq}
\end{equation}
where \(A_{W}\) and \(A_{H}\) are the respective loop functions for the SM
and charged-Higgs-induced contributions.  \(A_{H}\) is a  function of
\(m_{H^{\pm}}\) that has opposite sign to that of $A_W$. When
\(H^\pm\) is light, the charged Higgs
contribution to \(b\rightarrow s\gamma\) can be large and \(\sin\beta\) is either constrained to be
close to one, or $\sin\beta$ is tuned to a smaller value such that
$A_{W}+\cot^2\beta A_{H}\simeq -A_W$, thereby leading to an acceptable prediction.
If $m_{H^\pm}\gsim$ few TeV $A_H$ is effectively zero for any value of $\sin\beta\gsim 1/3$.
In both of these cases (light or heavy $H^\pm$) a solution exists for our purposes,
and the $H^\pm$ does not substantively affect the two-photon decay rate of the Higgs boson.
We specify \(\alpha\) to be small (in order to be
precise in our discussion, we choose \(\alpha=0\)), which is also nicely consistent
with $m_A^2,m_{H^0}^2,m_{H^\pm}^2\gg m_{h^0}^2$ and
the model contains only one light Higgs boson.  An extremely small \(\alpha\) is by no means
required, however, and we note that the results that carry forward will exhibit the same
qualitative features when $|\alpha|\lesssim1/2$.

\begin{figure}
\includegraphics[width=8.5cm]{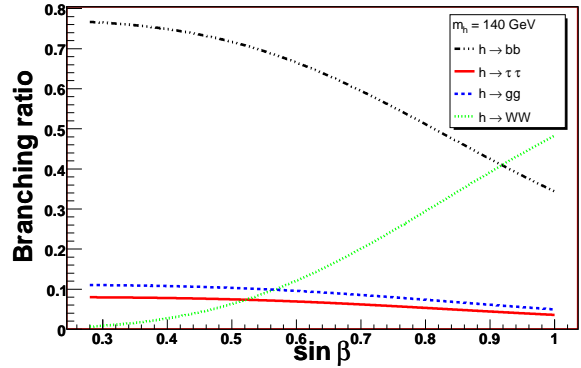}
\caption{Ratio with respect to the SM of decay branching fractions as a function of
$\sin\beta$ for the light Higgs of a type I two-Higgs doublet model. Here, we have taken
\(m_h=140\)~GeV and \(\alpha=0\).} \label{BRratio}
\end{figure}

{\it Numerical results.} In fig.~\ref{BRratio} we compute the decay branching fractions
of a $140\gev$ Higgs boson as a function of $\sin\beta$ relative to the SM decay
branching fractions.  In this, and subsequent calculations, SM quantities are
obtained using HDECAY~\cite{Djouadi:2000gu}, and we allow \(\sin\beta\) to vary from 1
down to \(\sim0.250\), below which point the perturbativity of the top quark Yukawa
(given by \(y_{t}=\sqrt{2}m_t/v\sin{\beta}\) in this model) becomes a concern.  For low
values of $\sin\beta$ the branching fraction to $b\bar b$ is enhanced significantly over
that of $WW^*$. As $\sin\beta\to 1$, which recovers the SM result, $WW^*$ wins out.

More important than the branching fractions, however, is the total cross-section of
$pp\to h\to\gamma\gamma$, since that is what is measured at the collider. The largest
contribution to the production cross-section for this observable $\sigma_h(\gamma\gamma)$
is through gluon fusion, $gg\to h\to \gamma\gamma$. The amplitude for $gg\to h$ is the
same as for $h\to gg$ up to simple co-factors at leading order. Thus, we make a good
estimate of the relative size of the $pp\to h\to\gamma\gamma$ cross-section: \beq
\frac{\sigma_h(\gamma\gamma)}{\sigma_h(\gamma\gamma)_{sm}} =
\frac{\Gamma_h(gg)}{\Gamma_h(gg)_{sm}}
\frac{\Gamma_h(\gamma\gamma)}{\Gamma_h(\gamma\gamma)_{sm}} \left( \frac{\Gamma_h({\rm
tot})}{\Gamma_h({\rm tot})_{sm}}\right)^{-1} \eeq

\begin{figure}
\includegraphics[width=8.5cm]{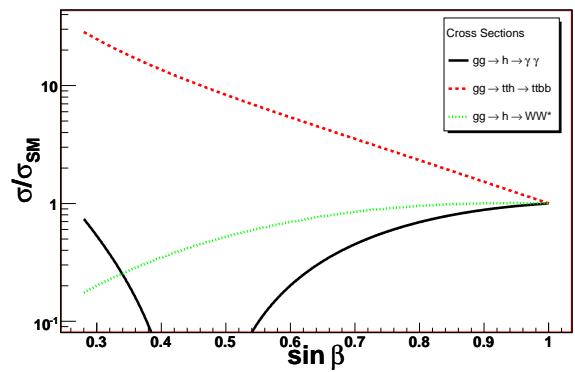}
\caption{Ratio with respect to the SM of cross-section observables at the LHC
(\(\sqrt{s}=14\)~TeV) as a function of $\sin\beta$ for the light Higgs of a type I
two-Higgs doublet model, with \(m_h=140\)~GeV and \(\alpha=0\).} \label{logobservables}
\end{figure}

In fig.~\ref{logobservables} we plot $\sigma_h(\gamma\gamma)/\sigma_h(\gamma\gamma)_{sm}$
as a function of $\sin\beta$ for $m_h=140\gev$ (and we retain $\alpha=0$). For
$0.38\lesssim \sin\beta \lsim 0.55$ we see that the total cross-section into two photons
is more than an order of magnitude lower than the SM cross-section, thereby severely
compromising the ability of the LHC to discover the light Higgs boson via that channel.
However, as we suggested above, the $t\bar t h\to t\bar t b\bar b$ cross-section is
greatly enhanced by nearly an order of magnitude above the SM cross-section.  Thus, it
may still be possible to discover such a Higgs boson.

\begin{table}
  \begin{center}
\begin{tabular}{|c|c|c|}\hline \(m_h\) (GeV) & \(\Gamma_{\mathit{tot}}(h)\) (MeV) &
   \(\Gamma_{\mathit{tot}}^{\mathit{sm}}(h)\) (MeV) \\ \hline
     110 & 12 & 3.0 \\
     130 & 16 & 4.9\\
     150 & 23 & 17\\ \hline
\end{tabular}
\end{center}
\caption{In this table, we list the total width of the Higgs boson, evaluated when the
\(h\gamma\gamma\) effective coupling is adjusted to zero, for a variety of choices of the
Higgs mass.  The SM width is also shown for
comparison\label{tab:HiggsWidth}.}
\end{table}

\begin{figure}
\includegraphics[width=8.5cm]{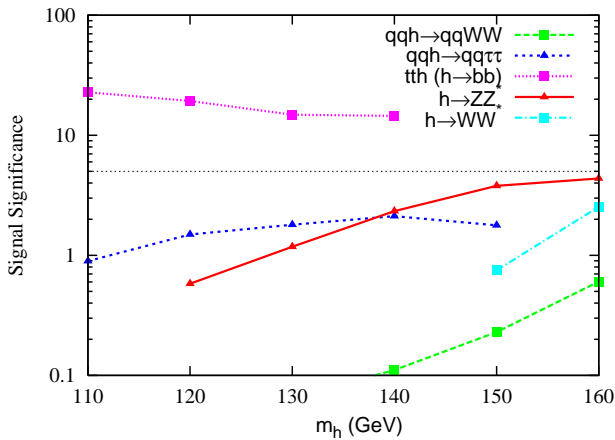}
\caption{Significance of various light Higgs boson observables at the LHC with
$30\xfb^{-1}$ of integrated luminosity. This plot is made for $\alpha=0$ and
$\sin\beta\simeq 0.45$ where $\Gamma(h\to\gamma\gamma)\to 0$ in the type I two-Higgs
doublet model.} \label{higgs significance}
\end{figure}

To investigate this, we compute the significance of discovery, with
\(\mathrm{30~fb}^{-1}\) of integrated luminosity, of the $h^0$ Higgs boson in this model
by scaling the known significance values~\cite{Asai:2004ws} for discovering a SM Higgs
boson in various channels using the ATLAS detector.
In fig.~\ref{higgs significance} we assume that $\alpha=0$ and
as we vary the Higgs boson mass we select the value of $\sin\beta$ such that $\Gamma(h\to
\gamma\gamma)$, and hence $\sigma(gg\to h\to \gamma\gamma)$, is zero.  Any value of
$\sin\beta$ in the neighborhood of this point will yield similar significance of
discovery values for the channels displayed.  The other significance values are modified
from their SM values by the scaling of the Higgs couplings by \(\eta_{W,Z}\) and
\(\eta_{f}\) according to equation~(\ref{eq:etas}).  Our scaling of the significance
values is justified since the Higgs boson width remains narrow and
never exceeds the invariant
mass resolutions (e.g., $\Delta m_{\gamma\gamma}/m_{\gamma\gamma}\sim 1\%$~\cite{ATLAStdr})
of the relevant final states over the range of interest -- see table~\ref{tab:HiggsWidth}.

Higgs boson discovery channels involving one or more
couplings between the Higgs and EW gauge bosons (including all weak boson fusion
processes) are suppressed by factors of \(\eta_{W,Z}\) as well as by an increase in the
total width of the Higgs due to an increase in Higgs decays to \(\overline{b}b\), which
dominate the total width in the mass range of interest---see table~\ref{tab:HiggsWidth}.
In contrast, the \(\overline{t} th\) channel significance is dramatically increased.  As
a result, while the usual channels in which one would look to discover a Higgs boson in
the mass range \(115\mathrm{~GeV}\lesssim m_{h}\lesssim 150\)~GeV are suppressed below
the \(5\sigma\) level, there is still the opportunity to discover the Higgs boson in the
$t\bar th$ channel. The searches, of course, must extend themselves to Higgs boson mass
values well above what is normally thought to be the relevant range for this signature.

{\it Conclusions.} Higgs decays to two photons provide one of the most important channels
through which one might hope to discover a light (\(115\mathrm{~GeV}\lesssim
m_{h}\lesssim 150\)~GeV) SM Higgs boson at the LHC. However, nature allows
a Higgs sector whose couplings differ from their SM values in such a way that
\(\Gamma(h\rightarrow\gamma\gamma)\) is suppressed into irrelevance.  This possibility
occurs readily in scenarios involving additional contributions to electroweak symmetry
breaking from sources that do not contribute to the generation of fermion masses. The
Higgs coupling to EW gauge bosons is decreased relative to the SM result as its
contribution to EWSB is decreased, and its coupling to the fermions is consequently
augmented, allowing for a cancelation in the coefficient of the effective
\(h\gamma\gamma\) vertex.  This is well illustrated
in type~I two Higgs doublet models and in a
variety of models where there exists a dynamical contribution to EWSB. If this is the
case, one cannot rely on the usual channels (weak boson fusion, \(h\rightarrow
WW^{\ast}\), and \(h\rightarrow\gamma\gamma\) itself) to discover a Higgs boson.
Nevertheless, such a Higgs could still be detected through processes like
\(\overline{t}th\), with \(h\) decaying to \(\overline{b}b\) or \(\overline{\tau}\tau\).




\end{document}